\documentclass[Review,sagev,times]{sagej}

\usepackage{moreverb,url}

\usepackage[colorlinks,bookmarksopen,bookmarksnumbered,citecolor=red,urlcolor=red]{hyperref}

\usepackage{soul}
\usepackage{xcolor}
\usepackage{setspace}
\usepackage{multirow}
\usepackage{natbib}
\usepackage{rotating}
\usepackage{ulem}
\begin{document}

\title{An Efficient Approach for Optimizing the Cost-effective Individualized Treatment Rule Using Conditional Random Forest}

\author{Yizhe Xu\affilnum{1}, Tom H. Greene\affilnum{1}, Adam P. Bress\affilnum{1}, Brandon K. Bellows\affilnum{2}, Yue Zhang\affilnum{1}, Zugui Zhang\affilnum{3}, Paul Kolm\affilnum{4},
 William S. Weintraub\affilnum{4}, Andrew S. Moran\affilnum{2}, Jincheng Shen\affilnum{1}}
\affiliation{\affilnum{1}Department of Population Health Sciences, University of Utah, SLC, Utah, US\\
\affilnum{2}Columbia University Medical Center, New York, NY, US\\
\affilnum{3}Christiana Care Health System, Newark, DE, US\\
\affilnum{4}Department of Medicine, MedStar Health Research Institute, Washington DC, DC, US}

\corrauth{Jincheng Shen, Department of Population Health Sciences, University of Utah, SLC, Utah 84108, U.S.}
\email{jincheng.shen@hsc.utah.edu}

\begin{abstract}
Evidence from observational studies has become increasingly important for supporting healthcare policy making via cost-effectiveness (CE) analyses. Similar as in comparative effectiveness studies, health economic evaluations that consider subject-level heterogeneity produce individualized treatment rules (ITRs) that are often more cost-effective than one-size-fits-all treatment. Thus, it is of great interest to develop statistical tools for learning such a cost-effective ITR (CE-ITR) under the causal inference framework that allows proper handling of potential confounding and can be applied to both trials and observational studies. In this paper, we use the concept of net-monetary-benefit (NMB) to assess the trade-off between health benefits and related costs. We estimate CE-ITR as a function of patients' characteristics that, when implemented, optimizes the allocation of limited healthcare resources by maximizing health gains while minimizing treatment-related costs. We employ the conditional random forest approach and identify the optimal CE-ITR using NMB-based classification algorithms, where two partitioned estimators are proposed for the subject-specific weights to effectively incorporate information from censored individuals. We conduct simulation studies to evaluate the performance of our proposals. We apply our top-performing algorithm to the NIH-funded Systolic Blood Pressure Intervention Trial (SPRINT) to illustrate the CE gains of assigning customized intensive blood pressure therapy.
\end{abstract}


\maketitle

\section{Introduction}\label{Intro}
Identifying treatment with improved efficacy is one of the central tasks in clinical research. Yet, effectiveness may not be the only consideration in practice. For example, treatments are sometimes found to have similar health benefits but very different treatment-related risks and costs. The randomized Prostate Cancer Intervention Versus Observation Trial (PIVOT) followed men with localized prostate cancer for nearly 20 years and found that prostatectomy was not associated with significantly lower all-cause or prostate-cancer mortality than expectant management\cite{b14}. However, prostatectomy costs three times as high as expectant management and induces a higher risk of complications. When two treatment have similar clinical benefits, healthcare policy makers often focus on comparing their costs as the practical and ultimate treatment allocation plan is the one that optimizes the utility of limited healthcare resources. In health economic research, treatments are evaluated using cost-effectiveness (CE) analysis in terms of incremental CE, which is usually measured as the contrast between health gains and additional treatment-induced costs.

Conventional CE analysis focuses on informing the best treatment plan at a population level using average incremental CE; however, this often leads to the ``one-size-fits-all'' treatment strategy that may not be optimal for patients whose responses to treatment vary from the population average. For instance, the National Lung Screening Trial\cite{b31} showed that the CE of screening with low-dose computed tomography varies by age, sex, smoking status, and risk of lung cancer. Analogous to the fact that individuals tend to have heterogeneous treatment benefits, a treatment may also induce different amount of cost for subjects with varying health status, e.g., additional medical procedures for managing complications\cite{b34}. All these sources of heterogeneity contribute to the variation in incremental CE among patients, which should be carefully considered to provide unique insights on personalized CE-based decision-making. Thus, our work estimates the optimal cost-effective individualized treatment rule (CE-ITR) that tailors treatment recommendations to subjects' characteristics and maximizes mean cost-effectiveness when applied for the entire population.

When the objective is to maximize a single health outcome, the optimal ITR can be estimated using standard model-based approaches by inverting the estimates of treatment-covariate interaction terms. However, such methods are vulnerable to model misspecification and they are developed to minimize prediction errors that do not necessarily lead to optimal regimes\cite{b7}. These concerns become more pressing when the goal is to maximize health gains while minimizing medical expenses. Even when both health and cost outcome models are correctly specified, they are likely to suggest different optimal rules that one may not know how to synchronize.  In contrast, classification based machine learning methods provide direct estimation of the optimal ITR while imposing fewer constraints on models. These advantages are extremely beneficial to CE studies where multiple outcomes are involved. So far, several different formulations have been proposed in this direction to recast the ITR estimation problem into a classification problem, and a variety of classification tools have been implemented to find the optimal ITR as the classifier that minimizes a suitably weighted misclassification rate\cite{b8,b9,b38}. Yet, the referenced methods are developed based on a single health outcome, so we aim to extend them to incorporate CE outcomes for health economic evaluations.

The determination of the best CE-ITR involves understanding both the treatment efficacy and differential costs at a patient level. As clinical research often studies efficacy using a time-to-event endpoint that is subject to censoring, we follow the biomedical literature\cite{b35} and focus on the restricted survival time measure that is more sensible for chronic disease studies with limited follow-up times. The treatment effect on cost is the difference in cumulative costs over a follow-up period. The estimation of cost has two major challenges: 1) positively skewed distribution\cite{b115}: Some prior work assumed the sample means of cost are normally distributed, which requires a sufficiently large sample size that may not be satisfied in some studies\cite{b45}; 2) induced informative censoring. Lin et al.\cite{b22} has pointed out that there is a positive correlation between cumulative costs at death and censoring, so standard survival models may be inappropriate for estimating total costs. Inaccurate estimation of cost may result in substantial bias in both model-based and classification-based methods and mislead healthcare policy making; thus, more flexible approaches should be applied to resolve these difficulties.

Claims data is one important resource of patient-level cost data, but when it is unavailable, researchers often obtain cost from microsimulations that are also feasible for randomized trials. A majority of CE analyses are conducted using microsimulation models (e.g., national population simulation models) to help quantify short- and long-term ramifications of healthcare decisions\cite{b4}. Some microsimulation-based CE analyses are free of censoring issues under certain model assumptions, while others are performed alongside clinical trials\cite{b55} where non-random loss-to-follow-up may occur. There are also CE studies that employ a combination of trial and observational data\cite{b56}, in which case the treatment-outcome confounding and censoring issues may co-exist. As healthcare policy makers increasingly demand real-world evidence of the economic value of an intervention, observational data will be widely used in the CE research in the near future. Therefore, robust statistical methods are desirable to accommodate a widespread nature of data.

Motivated by the \textit{net-monetary-benefit} (NMB) concept in CE analyses, our earlier work recommended to estimate the best CE-ITR by modeling patients' survival and cumulative cost simultaneously\cite{b60}. However, in practice, this method may suffer unstable estimation or loss in efficiency when the true CE-ITR has a complicated form or the censoring rate is extremely high, respectively. In this paper, we follow the composite outcome framework in our previous work and propose two improved approaches that account for confounding and induced informative censoring issues, with two layers of novelty. First, we propose a partitioned inverse-probability-weighting (IPW) estimator and a partitioned augmented IPW (AIPW) estimator for the NMB-based weights that are used in the classification step for identifying the optimal CE-ITR. The partitioned estimators more effectively incorporate available cost information from observed data therefore improve estimation efficiency; besides, they are robust to the skewed distribution of cost and  model misspecification. Second, we combine our weight estimators with an advanced statistical learning approach that is more suitable for modeling complex CE-ITRs and more effective in mitigating the common overfitting issue in machine learning and handling the correlated tailoring variables in the rule. 

The remainder of the paper is organized as follows. We introduce proposed methods in Section \ref{sec:method} and evaluate their performance as compared to several existing approaches via simulation studies in Section \ref{Sim}. In Section \ref{App}, we illustrate the top-performing approach from our simulations by evaluating the projected 15-year personalized CE of the intensive blood pressure therapy of the NIH-funded Systolic Blood Pressure Intervention Trial (SPRINT) study. Finally, in Section \ref{sec:diss}, we discuss the practical challenges in CE analysis and the potential extensions of our methods. 

\section{Method}\label{sec:method}

\subsection{Notations and Assumptions}
We consider an observational study with a two-level treatment $A$, where $A\in\{0,1\}$, so its distribution is driven by some of the baseline covariates $\mathbf{X}$. We denote the survival time by $T^{*}$ and censoring time by $C$, then the observed survival time $U^{*}=\mathrm{min}(T^{*},C)$. To incorporate the practical issue of limited follow-up in most studies, we define the health outcome as the restricted survival time, i.e., $U=\mathrm{min}(U^{*},\tau)$, where $\tau$ is the predetermined study time horizon of interest. We denote $M(t)$ as the accumulated cost up to an arbitrary time $t$, then, our cost outcome is the cumulative cost up to $U$, $M(U)$. Finally, we define the composite NMB outcome $Y=\lambda{U}-M(U)$, which scales health outcomes and the related cost in monetary value using the willingness-to-pay (WTP) parameter $\lambda$ that symbolizes the most one would like to spend for gaining another life-year. Thus, our observed data is  $O_{i}=\{X_{i},A_{i},C_{i},U_{i},M_{i}(U),Y_{i}\}_{i=1}^{n}$.

To simplify the discussion on heterogeneity in treatment effects and facilitate the illustration of our proposed method, we adopt the counterfactual framework in causal inference. Let $T^{*(a)}$ denote the counterfactual survival time under treatment $a$, then the potential restricted survival time $T^{(a)}=\mathrm{min}(T^{*(a)},\tau)$ and the treatment effect $\Delta{T}= T^{(1)}-T^{(0)}$. Analogously, $M^{(a)}(t)$ is the counterfactual cumulative cost up to time $t$ and the treatment-related incremental cost $\Delta{M}=M^{(1)}(T^{(1)})-M^{(0)}(T^{(0)})$. Note that the treatment effects on cost may vary across individuals under two circumstances: One is when subjects have varied  $T^{(a)}$, and the other is when subjects have similar $T^{(a)}$ but varied comorbidity status that may accumulate cost in various speeds. For simplicity, we suppress the time dependency part of $M^{(a)}(t)$ as $M^{(a)}$ in the following text when no confusion exists. We denote the counterfactual NMB as $Y^{(a)}$, then the treatment effect on NMB (i.e., incremental NMB) $\Delta{Y}=Y^{(1)}-Y^{(0)}$. To identify the counterfactual quantities related to treatment effects in NMB, we extend the classic identification assumptions in causal inference to this setting as the following\cite{b1}:
\begin{enumerate}
  \item Consistency: $T = AT^{(1)}+(1-A)T^{(0)}$ and $M = AM^{(1)}+(1-A)M^{(0)}$;
  \item Positivity: $P(A=a|X)>\epsilon$ for some positive value of $\epsilon$ and for $a\in\{0,1\}$;
  \item Ignorability: $T^{(a)}\perp A|X$ and $M^{(a)}\perp A|X$, for $a\in\{0,1\}$.
\end{enumerate}
Furthermore, we assume conditional non-informative censoring for survival time; i.e., $T\perp{C}|\{X, A\}$.

\subsection{Direct Estimation of the Optimal CE-ITR}
We denote an arbitrary ITR as a function of subjects characteristics $g(X)$. Let $Y^{g(X)}$ be the potential NMB when treatment is assigned as the rule in $g(X)$. We define the optimal CE-ITR $g^{\mathrm{opt}}(X)$ as the treatment rule that maximizes the average NMB; i.e., $g^{\mathrm{opt}}(X) \equiv \mathrm{arg}\max_{g\in\mathcal{G}}\,\mathbb{E}[Y^{g(X)}]$. Zhang et al.\cite{b8} proposed a classification framework for estimating ITR and showed that $g^{\mathrm{opt}}$ is also the optimal classifier that minimizes the weighted misclassification error. We extend this idea to our CE setting as
\begin{eqnarray}
g^{\mathrm{opt}}(X)=\mathrm{arg}\min_{g\in\mathcal{G}}\,\mathbb{E}\left[|W|I\{g(X)\ne{Z}\}\right], \label{eq:obj}\\
\mathrm{where}\,Z=I\{\Delta{Y}(X)>0\},\,\,\,W=\Delta{Y}(X).  \nonumber
\end{eqnarray}
Eq(\ref{eq:obj}) reformulates the estimation of the optimal CE-ITR problem as a weighted classification problem, in which $g(X)$ classifies subjects into two classes that defined by $Z=I\{\Delta{Y}(X)>0\}$. The class $Z=1$ includes patients for whom the treatment is cost-effective; so, the treatment should be assigned from a CE perspective. Conversely, since the incremental CE for patients in the class $Z=0$ is non-positive, the baseline treatment or control should be assigned as their cost-effective option. We refer to $Z$ as the \textit{class label} and $|W|$ as the \textit{classification weight}, which are both functions of $\Delta{Y}(X)$; thus, our classification algorithm considers cost and effectiveness jointly. We see no misclassification error is yielded when $g(X)$ is identical to the underlying truth of CE-ITR $Z$. However, when $g(X)\ne{Z}$, a penalty is induced by patients who are failed to receive their cost-effective interventions. Note that the classification weight $|W|$ is defined at an individual level; so, misclassifying subjects with large incremental NMBs will result in large penalties. The formulation in Eq(\ref{eq:obj}) is not unique, and alternative objective functions have been proposed by defining $Z$ and $W$ with other transformations of an outcome\cite{b9}.

Since CE analysis adopts a composite outcome, the relationship between baseline covariates and NMB is more complicated than typical comparative effectiveness studies that only consider a single health outcome. To capture the heterogeneity in incremental NMB, model-based methods usually specify complicated forms to include all possible treatment-covariate interaction terms. However, overly sophisticated models may mistake some of the noise for real signals and lead to overfitting, and excessively parsimonious models may be vulnerable to model misspecification. In contrast, the classification framework estimates ITRs in a data-driven fashion, so it has significant advantages in CE analyses where the underlying data structure is complex and nonlinear. 

Next, we propose a two-step procedure that directly identifies the optimal CE-ITR. We first describe the use of statistical learning methods in the process of identifying the optimal CE-ITR regarding solving the weighted classification problem in Eq(\ref{eq:obj}). Similar as in ITR literature, a broad spectrum of machine learning methods can be applied in our proposal for CE-ITR estimation. We first review the most common and popular choice, decision tree, and summarize their limitations as compared to forest-based methods. Then, we discuss the major concerns of employing classic random forest algorithm and introduce the advanced conditional forest as our classifier. After that, we discuss the estimation of classification weights in Eq(1) where we propose two partitioned estimators to efficiently estimate the NMB-based weights under a survival setting.   

\subsubsection{Decision Tree versus Random Forest Classification Methods}
Tree based algorithm is a common choice in decision making studies\cite{b100, b103, b108, b60}. A decision tree\cite{b12} is built using greedy algorithms in which the best split-point at each node is chosen by searching through all available features to minimize misclassification error. Also, its hierarchical structure naturally takes into account the interaction terms between predictors\cite{b21}. Since a tree algorithm puts more emphasis on subjects with larger treatment effects on NMB, individuals with small incremental NMB are more likely to be misclassified when the tree size is small, yet, fully grown trees may suffer from overfitting. Pruning is a technique in machine learning that searches algorithms to balance the complexity and the predictive accuracy of a tree. In our implementation using the R package \textbf{rpart}, we prune off any split that does not reduce the overall lack-of-fit by a certain amount, which is quantified using the \textit{complexity} parameter (cp). We choose the cp that produces the smallest 10-fold cross-validation error. Decision trees are favored for properties such as interpretability, transparency, and straightforward implementation. The major drawbacks are overfitting and instability\cite{b62}, which may result in estimators with low-bias and high-variances (``weak-learners").

A random forest\cite{b40}, as an ensemble approach, is a collection of independent and identical single-tree models, in which each tree casts a vote for the predicted class, and the class with the majority of the votes is the prediction of the forest. Compared to decision trees, random forests enhance prediction performance by reducing variances in two ways. First, bagging is a machine learning ensemble meta-algorithm designed to improve the stability and accuracy by averaging many weak learners, such as decision trees. Second, random feature selection, on top of bagging, reduces variance by weakening the correlations between trees. Unlike decision trees, random forests pick the best splitting variables among randomly sampled $m$ variables instead of looking through all $p$ covariates ($m<p$). The two techniques allow random forests to attenuate the overfitting issue and provide more generalizable models. Thus, forest-based algorithms have been widely used in decision making research such as identifying subjects with high disease risks in medicine \citep{Wongvibulsin2020} and making advertising plans \citep{Radcliffe07} or predicting stock market prices in marketing \citep{BASAK2019552}.

In CE-ITR estimation, it is crucial to identify the key tailoring variables from a potentially large pool of clinical characteristics and patient demographics. As the interpretation of tree models highly depends on selected variables, we avoid the biased feature selection issue by applying the conditional forest approach \cite{b52}. To deal with various types of covariates, the conditional forest conducts variable selection and variable splitting in two separate steps. In the first step, a global null hypothesis test is executed to evaluate the independence between candidate variables and outcome $Y$ and identify the variable $X$ with the strongest association to $Y$. For hypothesis testing on a particular variable $X$, conditional forest employs a permutation test framework and constructs a linear test statistic using the mean and covariance when conditional on all the other covariates and all possible permutations of $Y$. In the second step, the optimal split is chosen for $X$ to maximize a two-sample linear statistic measuring the discrepancy between two daughter nodes. Thus, conditional forest conducts a conditional permutation scheme\cite{b63} and provides unbiased variable selection even when predictors are in multiple types and correlated.

Permutation based importance of $X$ is typically computed as the decrease in prediction accuracy before and after shuffling the variable $X$. To reflect the true importance of each variable, conditional forest computes the \textit{conditional permutation based importance} in which the values of $X$ are shuffled within groups of correlated covariates $Z$\cite{Strobl08}. This is equivalent to conditional on $Z$, which resolves the confounding issue in standard random forest. In practice, one may determine the correlated covariate set $Z$ based on empirical association tests or experts' domain knowledge. As $Z$ may contain variables of different types, conditional forest defines a grid using all the binary cut-off points for splitting $X$ in the current tree and permutes $X$ within this grid. Note that the conditional grid varies across trees as each tree is built using different subsamples and randomly selected covariates. Besides, the grid complexity depends on pre-specified control parameters such as the depth of the tree and the number of trees. The final conditional importance is the aggregated mean over all trees.

In this paper, we implement the conditional forest classification using the R package \textbf{party}. We pre-prune each tree by setting the maximum depth to five and pick the number of candidate features at each node through 10-fold cross-validation. We further illustrate the performance of the conditional forest under our proposal in the SPRINT example where various types of predictors are used. Even though the conditional forest has shown superior performance than a decision tree, it has other limitations including higher model complexity, lower interpretability, and longer computation time. So, we build each forest with 50 trees in the simulation study and with 500 trees in the real data analysis where more covariates are used.

Although our proposal is implemented via a conditional random forest approach, many more sophisticated statistical learning methods, such as convolutional neural network and gradient boosting, can also be applied to perform weighted classification and serve as the underlying algorithm for our proposed method. A neural network models a binary outcome as a function of the linear combination of input variables, in which the cross-entropy is minimized via back-propagation. Gradient boosting machine and Bayesian additive regression tree model outcomes by iteratively fitting the residuals at each step. Although these more complicated machine learning algorithms presumably perform well even in extreme cases, such as large and complex data, our proposal using conditional forest presents a compromise between model interpretability and complexity.

\subsection{Efficient Estimation of Classification Weights Using Partitioned Estimators}\label{ssec:est}

\subsubsection{Review of the Existing Estimators}
In Eq(\ref{eq:obj}), both the class label and the classification weight are defined in terms of $\Delta{Y}$, which is a counterfactual quantity that needs to be estimated using observed data. A natural idea for estimating weights is applying regression models for survival time and cost outcomes separately, in which the treatment effect heterogeneity is modeled using treatment by covariate interaction terms. This weight proposal is referred to as the \textit{Reg-based} estimator\cite{b60}, where the treatment
effect on restricted survival time is estimated as $\Delta{\hat{T}_{i}}=\int_{0}^{\tau}\{\hat{S}^{(1)}(t|X_{i};\hat{\alpha})-\hat{S}^{(0)}(t|X_{i};\hat{\alpha})\}dt$ and the survival function $S^{(a)}$ is estimated using a parametric survival model; The incremental cost is estimated using a generalized linear model with a gamma distribution as $\Delta{\hat{M}_{i}}=\mathbb{E}[\hat{M}^{(1)}_{i}-\hat{M}^{(0)}_{i}|X_{i};\hat{\beta}]$. Finally, $\hat{W}_{i}^{\mathrm{Reg-based}}=\lambda\Delta{\hat{T}_{i}}-\Delta{\hat{M}_{i}}$. Note that $\hat{W}_{i}^{\mathrm{Reg-based}}$ is the estimated individual-level treatment effect on NMB; so, one may simply determine the optimal regime based on the sign of the weight; i.e., $\hat{g}^{\mathrm{Reg-naive}}=I\{\hat{W}_{i}^{\mathrm{Reg-based}}>0\}$\cite{b60}. \textit{Reg-naive} is a model-based approach since the estimation of the optimal ITR only involves inverting regression estimates without a classification step.

Xu et al.\cite{b60} proposed a non-partitioned AIPW estimator for the NMB weights, and we refer to it as \textit{AIPW-NP}. The AIPW-NP estimator employs inverse probability treatment weighting (IPCW) and censoring weighting (IPTW) to account for confounding and right censoring, respectively, as follows:
\begin{eqnarray*}
\Delta{\hat{M}_{i}^{\mathrm{AIPW-NP}}}&=& \left[\frac{A_{i}M_{i}\delta_{i}}{\hat{e}^{M}_{i}\hat{K}_{1}(U_{i})}-\frac{(A_{i}-\hat{e}^{M}_{i})m_{1}(X_{i};\hat{\beta})\delta_{i}}{\hat{e}^{M}_{i}\hat{K}_{1}(U_{i})}\right]\\
                   &-&\left[\frac{(1-A_{i})M_{i}\delta_{i}}{(1-\hat{e}^{M}_{i})\hat{K}_{0}(U_{i})}+\frac{(A_{i}-\hat{e}^{M}_{i})m_{0}(X_{i};\hat{\beta})\delta_{i}}{(1-\hat{e}^{M}_{i})\hat{K}_{0}(U_{i})}\right];\\
\Delta{\hat{T}_{i}^{\mathrm{AIPW-NP}}}&=& \left[\frac{A_{i}U_{i}\delta_{i}}{\hat{e}^{T}_{i}\hat{K}_{1}(U_{i})}-\frac{(A_{i}-\hat{e}^{T}_{i})h_{1}(X_{i};\hat{\alpha})\delta_{i}}{\hat{e}^{T}_{i}\hat{K}_{1}(U_{i})}\right]\\
                   &-&\left[\frac{(1-A_{i})U_{i}\delta_{i}}{(1-\hat{e}^{T}_{i})\hat{K}_{0}(U_{i})}+\frac{(A_{i}-\hat{e}^{T}_{i})h_{0}(X_{i};\hat{\alpha})\delta_{i}}{(1-\hat{e}^{T}_{i})\hat{K}_{0}(U_{i})}\right];\\
\hat{W}_{i}^{\mathrm{AIPW-NP}} &=& \lambda\times{\Delta{\hat{T}_{i}^{\mathrm{AIPW}}}}-\Delta{\hat{M}_{i}^{\mathrm{AIPW}}},
\end{eqnarray*}
where $\hat{e}^{T}_{i}$ and $\hat{e}^{M}_{i}$ are the estimated propensity scores (PSs) for the two different outcomes; $\delta_{i}$ is the event indicator and $\hat{K}_{A_{i}}(U_{i})=P(\delta_{i}=1|A_{i}, X_{i})$ is the censoring weight. $h_{a}(X_{i};\hat{\alpha})$ and $m_{a}(X_{i};\hat{\beta})$ are regression estimates of survival time and cost, respectively. Due to right censoring, the AIPW-NP estimator uses the uncensored subjects to represent censored ones, which is appropriate when only total cost data is available. However, many databases such as the Electronic Health Record and the Medical Expenditure Panel Survey also record cost history data (e.g., monthly cost), which can be used to improve the estimation of classification weights. In the following sections, we propose two partitioned estimators for the NMB-based weights that incorporate cost history information from censored subjects.

\subsubsection{The Partitioned IPW Estimator Using Cost History Data}
One limitation of the AIPW-NP estimator is information loss, meaning that it is incapable of using the incomplete outcome data of censored subjects. Thus, we follow the ideas of Bang and Tsiatis\cite{b10} and propose a \textit{partitioned} IPW (\textit{IPW-P}) estimator for the NMB-based classification weights using cost history data. 

We partition the entire study period $(0,\tau)$ into $J$ subintervals $(t_{j},t_{j+1}]$, where $0=t_{1}<t_{2}<...<t_{J}<t_{J+1}=\tau$. Let $M_{i}^{j}$ be the cost of subject $i$ accrued from the $j^{\mathrm{th}}$ subinterval, $U_{i}^{j}=\min(U_{i}^{t_{j}},C_{i})$, and $\delta_{i}^{j}=I(U_{i}^{t_{j}}\leq{C_{i}})$, where $U_{i}^{t_{j}}=\min(U_{i},t_{j})$. With these subinterval-specific data, we construct the IPW-P estimator for incremental cost $\Delta{M}_{i}$ as the sum of interval-level weights:
\begin{eqnarray*}
\Delta\hat{M}_{i}^{\mathrm{IPW-P}}=\sum_{j=1}^{J}\frac{A_{i}M_{i}^{j}\delta_{i}^{j}}{\hat{e}_{i}^{M}\hat{K}_{1}(U_{i}^{j})}
                                      -\frac{(1-A_{i})M_{i}^{j}\delta_{i}^{j}}{(1-\hat{e}_{i}^{M})\hat{K}_{0}(U_{i}^{j})},
\end{eqnarray*}
where $A_{i}$ and $\hat{e}_{i}^{M}$ remain the same across different subintervals as the treatment is fixed over time, and $\delta_{i}^{j}$ and $\hat{K}_{A_{i}}(U_{i}^{j})$ are the event indicator and censoring weight for the $j^{\mathrm{th}}$ interval, respectively. Recall a non-partitioned IPW estimator can only use the total cost of uncensored subjects and discard the incomplete cost data of censored subjects even though they are counted by uncensored participants with whom they share similar baseline characteristics. In contrast, our IPW-P approach improves the weight estimation in two aspects: 1) Maximizing the utilization of cost data. By partitioning the entire follow-up, IPW-P employs the cost data of censored subjects from the subintervals before getting censored, so it makes use of incomplete data and improves estimation efficiency. As IPW-P gains more information from a longer pre-censoring period, it shows a greater efficiency improvement when subjects are censored at a later time of the follow-up. 2) Use of subinterval-specific censoring weights. The IPW-P estimator allows using uncensored or ``not-yet" censored subjects to represent the censored ones in each interval, which provides a richer pool for finding better representers. For instance, suppose subjects A and B share the most similar traits and are censored in intervals five and eight, respectively, and subject C is uncensored. A non-partitioned estimator can only use subject C to represent subject A for all subintervals, even if subject B is a better match. While the IPW-P estimator can use subject B to represent subject A from subintervals five to seven and to identify the next closet match, uncensored or not-yet censored subject, for the next interval. This flexible re-weighting scheme allows censored individuals to be accounted by the most similar set of subjects available over time, which improves the matching quality and results in lower bias as compared to situations where only one uncensored subject can be used for representation. Note that we describe this bias reduction under a specific setting where the matching or representing process may be imperfect.

With sufficiently small subintervals, every censored subject may contribute one or more pre-censoring periods, which guarantees a meaningful classification weight for cost for every subject in the study sample. To be consistent, we construct an IPW-R estimator for restricted survival time in which the weights of censored subjects are estimated using regression as follows:
\begin{eqnarray*}
\Delta\hat{T}_{i}^{\mathrm{IPW-R}} &=& \delta_{i}\left[\frac{A_{i}U_{i}}{\hat{e}_{i}^{T}}-\frac{(1-A_{i})U_{i}}{(1-\hat{e}_{i}^{T})}\right]\\
&+&(1-\delta_{i})\left[A_{i}h_{1}(X_{i};\hat{\alpha})-(1-A_{i})h_{0}(X_{i};\hat{\alpha})\right]
\end{eqnarray*}
Thus, we estimate the NMB-based classification weight as $\hat{W}_{i}^{\mathrm{IPW-P}} = \lambda{\Delta\hat{T}_{i}^{\mathrm{IPW-R}}}-\Delta\hat{M}_{i}^{\mathrm{IPW-P}}$. With a slight abuse of notation, we consider the IPW-P weight of subject $i$ as an approximation of $\Delta{Y_{i}}$ since it asymptotically estimates the treatment effect on NMB\cite{b13}.

\subsubsection{The Partitioned AIPW Estimator Using Cost History Data}
A natural extension of the IPW-P estimator is a partitioned AIPW estimator. We follow the ideas of Li et al.\cite{b11} but propose the AIPW-P estimator in a disaggregated form for estimating the individual-level NMB-based weights as follows:
\begin{eqnarray*}
\Delta{\hat{M}_{i}^{\mathrm{AIPW-P}}}&=&\sum_{j=1}^{J}\left[\frac{A_{i}M_{i}^{j}\delta_{i}^{j}}{\hat{e}_{i}^{M}\hat{K}_{1}(U_{i}^{j})}
                                     -\frac{(A_{i}-\hat{e}_{i}^{M})m_{1}^{j}(X_{i};\hat{\beta})\delta_{i}^{j}}{\hat{e}_{i}^{M}\hat{K}_{1}(U_{i}^{j})}\right]\\
                                   &-&\left[\frac{(1-A_{i})M_{i}^{j}\delta_{i}^{j}}{(1-\hat{e}_{i}^{M})\hat{K}_{0}(U_{i}^{j})}+\frac{(A_{i}
                                     -\hat{e}_{i}^{M})m_{0}^{j}(X_{i};\hat{\beta})\delta_{i}^{j}}{(1-\hat{e}_{i}^{M})\hat{K}_{0}(U_{i}^{j})}\right],\\
\Delta{\hat{T}_{i}^{\mathrm{AIPW-R}}}&=&\delta_{i}\left[\frac{A_{i}U_{i}}{\hat{e}_{i}^{T}}-\frac{(A_{i}-\hat{e}_{i}^{T})h_{1}(X_{i};\hat{\alpha})}{\hat{e}_{i}^{T}}\right]\\
                                    &-&\delta_{i}\left[\frac{(1-A_{i})U_{i}}{(1-\hat{e}_{i}^{T})}+\frac{(A_{i}-\hat{e}_{i}^{T})h_{0}(X_{i};\hat{\alpha})}{(1-\hat{e}_{i}^{T})}\right]\\
                                     &+&(1-\delta_{i})\left[A_{i}h_{1}(X_{i};\hat{\alpha})-(1-A_{i})h_{0}(X_{i};\hat{\alpha})\right],\\
\hat{W}_{i}^{\mathrm{AIPW-P}} &=& \lambda{\Delta\hat{T}_{i}^{\mathrm{AIPW-R}}}-\Delta\hat{M}_{i}^{\mathrm{AIPW-P}}
\end{eqnarray*}
where $m_{a}^{j}(X_{i};\hat{\beta})$ is the regression estimate of the cost that accrued in the $j^{\mathrm{th}}$ subinterval. We say the AIPW-P estimator makes ``double information gain" as it incorporates extra information from outcome models that captures the relationship between covariates and outcomes when compared to IPW-P, and it utilizes most of the cost data from censored subjects as an additional information source when compared to AIPW-NP.

In principle, both ways of acquiring additional information can help improve estimation performance; though, several factors may influence the gain from these modifications. Censoring distribution and censoring rate may impact the amount of data that could be obtained from censored subjects. As mentioned earlier, the later the subjects are censored during a follow-up (e.g., censored at year 10 versus year 1), and the higher the censoring rate is (e.g., 50\% versus 10\%), the more cost data, i.e., more efficiency, may be gained by using a partitioned weight estimator. Since studies with an early censoring (e.g., one-year pre-censoring time) are likely to have high censoring rates (50\%), and studies with a late censoring (e.g., 10-year pre-censoring time) tend to have low censoring rates (10\%), the partitioned estimators gain either a small amount of data from many subjects or a large amount of data from a few patients. Furthermore, high censoring rates may affect the accuracy of survival time estimator used in the augmentation terms, e.g., due to extrapolation in parametric survival models\cite{b57}. An alternative is to apply nonparametric survival methods such as random survival forest or Bayesian additive regression trees. In addition, conventional regression has limited ability in modeling high-dimensional covariates, so one may apply regularized regression such as lasso or elastic-net, or nonparametric machine learning methods such as random survival forest or causal survival forest. 

To summarize, we proposed a two-step procedure: In the first step, we estimate the NMB-based individual weights using the IPW-P or AIPW-P estimators. In the second step, we plug these estimated weights into Eq(1) and solve it via the statistical learning algorithms of our choice to estimate the optimal CE-ITR. 

\section{Simulation Study} \label{Sim}

\subsection{Simulation Schemes}
In this section, we conduct simulation studies to evaluate the finite sample performance of our proposed methods. We consider five identical independently distributed covariates $\mathbf{X}$, where $\mathbf{\tilde{X}}= \{X_{1},X_{2}\}$ are  $N(1,2)$ and $X_{3}$, $X_{4}$ and $X_{5}$ are $N(0,1)$. We generate the treatment assignment $A$ using a logistic regression model as $\mathrm{logit}(A=1) = 0.5X_{1}+0.5X_{2}+0.9X_{3}$. We simulate the NMB outcome in two parts, the survival time and cumulative cost, and we consider two commonly used values in CE analyses, \$50,000 or \$100,000 per year of life, for the WTPs\cite{b41}.

We generate the potential survival times using an exponential proportional hazard model with a hazard rate of $0.1 \mathrm{exp} (\mathbf{X} \beta_{T} - \mathbf{\tilde{X}} \gamma_{T} A)$, where $\beta_{T}$ and $\gamma_{T}$ are the coefficients for main effect terms and interaction terms, respectively. Every subject has at least 5 years of follow-up. Note that $X_{1}$, $X_{2}$, and $X_{3}$ are confounders as they predict both treatment and survival time, and $X_{1}$ and $X_{2}$ are also effect modifiers. We let $\beta_{T} = (0.8, 0.8, 0.3, 0.3, 0.3)^{\mathrm{T}}$, $\gamma_{T}=(2.0, 1.5)^{\mathrm{T}}$ and $\gamma_{T}=(2.5, 2.0)^{\mathrm{T}}$ for small and large heterogeneous treatment effect (HTE), respectively. We consider a wide scope of censoring rate, ranging from complete observations to heavy censoring, by generating the censoring time using an exponential proportional hazard model with different baseline hazards. We choose the restriction time as $\tau=20$ years.

We follow a standard health economic simulation setting\cite{b11} and construct the cumulative cost with three components: an initial cost ($M_{I}$), an ongoing cost for every 6-month ($M_{O}$), and a death-related cost ($M_{D}$). All costs are generated from a Gamma distribution with varied scale parameters: $M_{I}^{(a)}\sim{\Gamma{(\kappa,\,\theta_{(a)})}},$ $M_{O}^{(a)}\sim{\Gamma{(\kappa,0.6\theta_{(a)})}},$ and $M_{D}^{(a)}\sim{\Gamma{(\kappa,0.2\theta_{(a)})}}$. We set $\kappa=2.5$ and $\theta_{(a)}=\mathrm{exp}\{\mathbf{X_{0}}\beta_{M}+\gamma_{M}(X_{1}+X_{2})a\}$ or $\mathrm{exp}(\mathbf{X_{0}}\beta_{M}+2\gamma_{M}a)$ for present or absent HTE on cost, respectively, where $\beta_{M}= (0.8, 0.8, 0.3, 0.3, 0.3)^{\mathrm{T}}$ and $\gamma_{M}=0.03$. The potential cumulative cost $M^{(a)}$ is simulated as $1500\{M_{I}^{(a)}+M_{O}^{(a)}T^{(a)}+M_{D}^{(a)}D^{(a)}\},$ where $D^{(a)}=I\{T^{*(a)}\leq{\tau}\}$. In observed data, we only count the death-related cost for subjects whose deaths are observed before the end of the study; so, $M_{i} = 1000\{M_{I}^{(A_{i})}+M_{O}^{(A_{i})}U_{i}+M_{D}^{(A_{i})}D_{i}\},$ where $D_{i}=I\{T_{i}^{*}\leq{\mathrm{min}(C_{i},\tau)}\}$.

\subsection{Simulation Scenarios}
We design 32 different simulation scenarios by varying four parameters: 1) Presence of treatment effect modification on both cost and survival times versus on survival times only, denoted as \textit{EM-TM} and \textit{EM-T}; 2) the magnitude of heterogeneity in treatment effects on the hazard ratio scale for restricted survival time, referred to as \textit{small HTE} and \textit{large HTE}, defined above; 3) two different WTP thresholds, \$50K and \$100K per year of life; 4) four different censoring rates, 0\%, 20\%, 50\%, 70\%. Under each scenario, we simulate 500 data sets, each with 1000 samples. Our entire simulation is conducted using parallel processing. 

We apply seven direct methods named \textit{DT-AIPW-NP}, \textit{DT-IPW-P}, \textit{DT-AIPW-P}, \textit{CRF-AIPW-NP}, \textit{CRF-IPW-P}, \textit{CRF-AIPW-P}, and \textit{Reg-naive} to estimate the optimal CE-ITR, in which CRF-IPW-P and CRF-AIPW-P are our proposed methods. For each name, the first segment indicates the classifier choice, a weighted decision tree (DT) or conditional random forest (CRF), and the second and third segments specify the type of classification weight estimators introduced in Section \ref{ssec:est}. To assess the impact of model misspecification on method performance, we correctly specify the treatment and censoring models but misspecify the survival time and cost outcome models by omitting the treatment-covariate interaction term $AX_{1}$.

We evaluate the performance of all seven methods for each of the 16 scenarios using two metrics. First, we compare the classification accuracy of an estimated optimal ITR $\hat{g}^{\mathrm{opt}}$, where the accuracy is defined as the proportion of subjects that are classified to the correct treatment group based on the true optimal ITR $g^{\mathrm{opt}}$. Second, we compute the mean NMB under each estimated optimal ITR as $\mathbb{E}[Y^{(1)}\hat{g}^{\mathrm{opt}}+Y^{(0)}(1-\hat{g}^{\mathrm{opt}})]$ and compare it to the mean NMB under the true optimal regime $\mathbb{E}[Y^{g^{\mathrm{opt}}}]$.

\subsection{Simulation Results}
Table \ref{t:CCR} shows the average classification accuracy under 16 different scenarios when WTP is \$50K. The standard error (SE) represents the variability of the estimates across 500 simulated data sets. Since both outcome models for survival times and cost are misspecified, the Reg-naive method yielded the least accurate results among all methods. Given a fixed classifier, methods that use partitioned AIPW based NMB weights resulted in higher accuracy than the ones using non-partitioned weights across all four levels of censoring. The partitioned IPW based weights outperformed and underperformed the non-partitioned estimators when the censoring is low ($\leq 20\%$) and high ($\ge 50\%$), respectively. On the other hand, with a fixed weight estimator, methods that apply a conditional forest classifier resulted in higher accuracy than the ones using a decision tree. Thus, the CRF-AIPW-P method, using both partitioned weights and conditional random forest classifiers, yielded the most accurate results among all methods and across all scenarios.

Reg-naive and methods that use non-partitioned weights (DT-AIPW-NP and CRF-AIPW-NP) showed lower CCRs at low censoring rates (0\% and 20\%) than at high censoring rates (50\% and 70\%). One possible factor is the misspecified outcome models used in the weight estimation step. In contrast, the IPW based approaches (DT-IPW-P and CRF-IPW-P), without using outcome models, showed a decreasing trend in CCR as the censoring rate goes up. With a high censoring rate, the pre-censoring time of censored subjects was reduced, so the IPW-P estimator had fewer subintervals to gain cost information. The AIPW-P based methods (DT-AIPW-P and CRF-AIPW-P) showed consistently accurate and efficient performance across different levels of censoring since they have two different ways to gain information and remain robust. Methods that use partitioned estimators showed improved performance when the HTE on survival times increased to a larger value, which amplifies the heterogeneity in incremental NMB and yields stronger signals for classifiers to detect. The results under scenarios \textit{EM-TM} and \textit{EM-T} have similar patterns. Results for WTP is \$100K are provided in Tables \ref{t:CCR100K} and \ref{t:EMO100K} in the Supplemental Materials.

\begin{sidewaystable}
\small
\centering
\caption{Comparison of Classification Accuracy (\%) of Estimated Optimal ITRs. ``EM=TM" indicates the presence of effect modification on both survival time and cost; ``EM=T" indicates the presence of effect modification on survival time only. ``CR" is short for ``censoring rate". ``HTE" indicates the amount of effect modification on the hazard ratio scale for survival time: ``HTE=S" for small effect modification and ``HTE=L" for large effect modification. Willingness-to-pay = \$50K.}
\label{t:CCR}
\begin{tabular}{lllccccccc}
\hline
&&&&&&&&&\\
EM& CR & HTE & \multicolumn{1}{c}{Reg-naive} & \multicolumn{1}{c}{DT-AIPW-NP} & \multicolumn{1}{c}{DT-IPW-P}& \multicolumn{1}{c}{DT-AIPW-P} & \multicolumn{1}{c}{CRF-AIPW-NP} & \multicolumn{1}{c}{CRF-IPW-P}& \multicolumn{1}{c}{CRF-AIPW-P}\\
\hline
\multirow{8}{*}{TM}& \multirow{2}{*}{0\%}  &S& 51.9 (3.5) & 75.6 (3.6) & 85.3 (1.4) & 86.6 (1.2) & 78.8 (3.3) & 86.3 (1.2) & 87.2 (1.2) \\
                   &                       &L& 45.8 (3.2) & 71.3 (4.3) & 86.9 (1.4) & 88.1 (1.1) & 73.4 (5.0) & 87.7 (1.2) & 88.4 (1.1) \\\cmidrule{2-10}
                   & \multirow{2}{*}{20\%} &S& 53.8 (3.3) & 76.6 (3.5) & 85.1 (1.4) & 86.4 (1.3) & 80.3 (3.3) & 86.3 (1.2) & 87.0 (1.2) \\ 
                   &                       &L& 47.3 (3.2) & 73.4 (3.9) & 86.9 (1.4) & 87.9 (1.2) & 76.3 (4.8) & 87.7 (1.2) & 88.3 (1.1) \\ \cmidrule{2-10}
                   & \multirow{2}{*}{50\%} &S& 72.4 (1.9) & 76.2 (3.7) & 60.7 (5.7) & 86.3 (1.4) & 80.6 (3.3) & 55.9 (7.7) & 86.8 (1.1) \\
                   &                       &L& 72.0 (2.1) & 77.0 (4.0) & 70.2 (5.2) & 87.8 (1.3) & 81.0 (3.3) & 73.9 (5.8) & 88.0 (1.2) \\ \cmidrule{2-10}
                   & \multirow{2}{*}{70\%} &S& 73.0 (1.4) & 71.4 (3.9) & 65.5 (3.4) & 82.9 (1.8) & 74.5 (4.4) & 68.6 (3.1) & 83.2 (1.6) \\ 
                   &                       &L& 74.2 (1.4) & 77.7 (3.3) & 63.4 (3.7) & 85.6 (1.6) & 81.5 (2.9) & 66.4 (3.4) & 85.5 (1.5) \\\hline
\multirow{8}{*}{T} & \multirow{2}{*}{0\%}  &S& 55.0 (3.8) & 77.2 (4.3) & 89.1 (1.3) & 90.0 (1.1) & 81.9 (3.9) & 90.2 (1.0) & 90.6 (1.0) \\
                   &                       &L& 46.5 (3.5) & 71.1 (4.9) & 90.5 (1.2) & 91.4 (1.0) & 75.6 (5.6) & 91.6 (1.0) & 91.8 (1.0) \\\cmidrule{2-10}
                   & \multirow{2}{*}{20\%} &S& 57.6 (3.6) & 78.8 (4.1) & 89.0 (1.3) & 90.0 (1.2) & 83.1 (3.7) & 90.2 (1.1) & 90.5 (1.1) \\ 
                   &                       &L& 48.8 (3.5) & 73.9 (4.6) & 90.5 (1.2) & 91.4 (1.0) & 78.6 (4.7) & 91.6 (1.0) & 91.8 (1.0) \\ \cmidrule{2-10}
                   & \multirow{2}{*}{50\%} &S& 76.6 (1.9) & 80.2 (4.1) & 64.9 (5.8) & 89.8 (1.2) & 85.2 (3.1) & 66.9 (7.3) & 90.2 (1.0) \\
                   &                       &L& 76.1 (2.1) & 80.8 (4.5) & 76.1 (5.0) & 91.2 (1.1) & 85.5 (3.3) & 82.5 (4.2) & 91.6 (1.0) \\ \cmidrule{2-10}
                   & \multirow{2}{*}{70\%} &S& 76.3 (1.3) & 74.9 (4.4) & 69.4 (3.1) & 86.7 (1.6) & 78.6 (4.4) & 73.2 (2.8) & 86.9 (1.5) \\ 
                   &                       &L& 77.5 (1.3) & 81.4 (3.4) & 67.9 (3.4) & 89.0 (1.6) & 85.3 (3.1) & 71.1 (3.3) & 89.0 (1.4) \\ \hline
\end{tabular}
\end{sidewaystable}

\begin{sidewaysfigure}
\centering
\centerline{\includegraphics[scale=0.43]{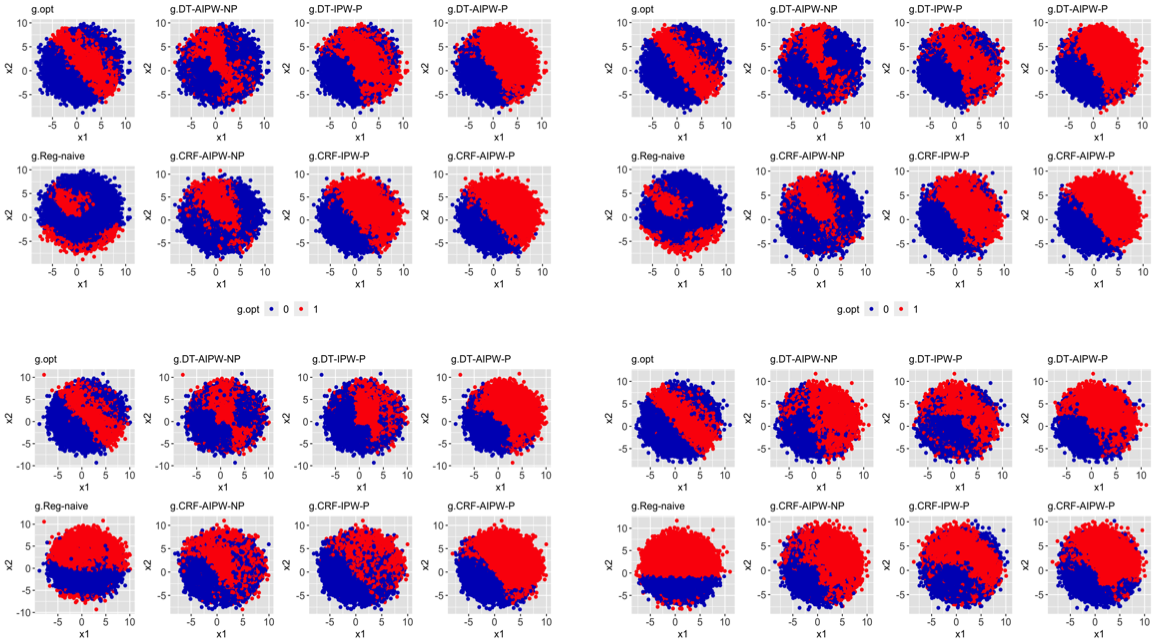}}
\caption{Comparison of Decision Boundaries of Estimated Optimal ITRs. Presence of effect modification on both survival time and cost, and only the outcome models are misspecified. WTP = \$50K and large HTE. The four panels are labeled as follows. Top left:, CR=0\%; Top right:, CR=20\%; Bottom left: CR=50\%; Bottom right: CR=70\%. Each panel contains eight figures: the very first figure shows the true decision boundary (gold standard), and the rest seven display the estimated decision boundaries that are given by seven different methods.}
\label{f:SimF3}
\end{sidewaysfigure}

\begin{sidewaysfigure}
\centering
\centerline{\includegraphics[scale=0.43]{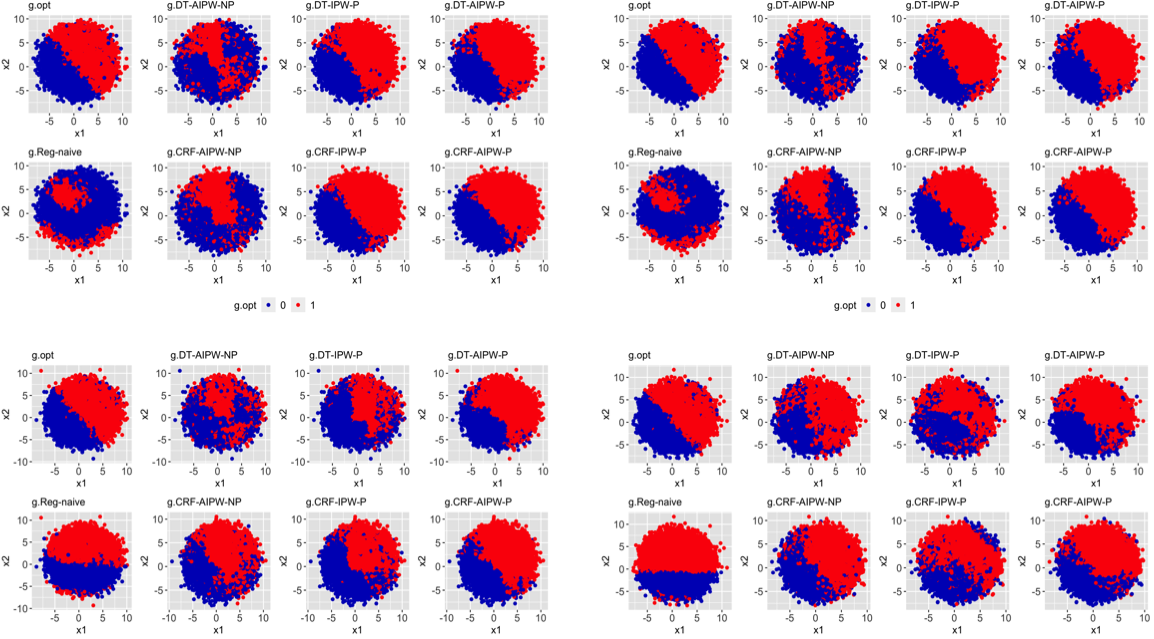}}
\caption{Comparison of the Decision Boundaries of the Estimated Optimal ITRs. Presence of effect modification on survival time but not on cost, and only the outcome models are misspecified. WTP = \$50K and large HTE. The four panels are labeled as follows. Top left:, CR=0\%; Top right:, CR=20\%; Bottom left: CR=50\%; Bottom right: CR=70\%. Each panel contains eight figures: the very first figure shows the true decision boundary (gold standard), and the rest seven display the estimated decision boundaries that are given by seven different methods.}
\label{f:SimF3a}
\end{sidewaysfigure}

Table \ref{t:EMO} shows that more accurate ITRs resulted in less biased mean NMBs as expected. Figures \ref{f:SimF3} and \ref{f:SimF3a} show that both partitioned estimators and the conditional random forest classifier help to improve the decision boundary estimation (true decision boundary is labeled as ``g.opt"). In Figure \ref{f:SimF3}, even though AIPW-P based methods did not capture the unclear decision boundary on the upper right corner, they outperformed all the other methods on estimating the clear decision on the lower left corner. In addition, the estimated decision boundaries deviate more from the truth as censoring rate increases. Among all seven methods, the CRF-AIPW-P yielded the most similar decision boundaries to the truth. The decision boundary graphs for other scenarios are provided in the Supplemental Materials. The implementation via R software of our proposed method is computationally efficient. It took on average approximately 2.7 sec for a typical analysis to estimate the optimal CE-ITR from a cohort with 1000 samples on an Intel\textsuperscript{®} Xeon\textsuperscript{®} E5-2667 computer.

\begin{sidewaystable}
\fontsize{8}{12}\selectfont
\centering
\caption{Comparison of Mean Outcomes under Different Estimated Optimal ITRs. The indications of ``EM", ``CR", ``WTP", and ``HTE" are the same as those in Table 1. The $g^{\mathrm{opt}}$ column represents the mean NMB (in the unit of 10,000) under the true optimal regime, and the other seven columns on the right display the mean NMB under the corresponding estimated regime. Willingness-to-pay = \$50K.}
\label{t:EMO}
\begin{tabular}{lllcccccccc}
\hline
&&&&&&&&&&\\
EM& CR & HTE &\multicolumn{1}{c}{$\mathbb{E}[Y^{g^{\mathrm{opt}}}]$}& \multicolumn{1}{c}{Reg-naive} & \multicolumn{1}{c}{DT-AIPW-NP} & \multicolumn{1}{c}{DT-IPW-P}& \multicolumn{1}{c}{DT-AIPW-P} & \multicolumn{1}{c}{CRF-AIPW-NP} & \multicolumn{1}{c}{CRF-IPW-P}& \multicolumn{1}{c}{CRF-AIPW-P}\\
\hline
\multirow{8}{*}{TM}& \multirow{2}{*}{0\%}  &S& 17.1 (0.4) & 9.7 (0.7) & 14.4 (0.7) & 16.0 (0.4) & 16.2 (0.4) & 14.8 (0.7) & 16.1 (0.4) & 16.3 (0.4) \\ 
                   &                       &L& 18.1 (0.4) & 9.1 (0.7) & 14.2 (0.9) & 17.0 (0.4) & 17.2 (0.4) & 14.4 (1.0) & 17.1 (0.4) & 17.3 (0.4) \\ \cmidrule{2-11}
                   & \multirow{2}{*}{20\%} &S& 17.1 (0.4) & 9.9 (0.6) & 14.5 (0.7) & 16.0 (0.4) & 16.2 (0.4) & 15.1 (0.7) & 16.2 (0.4) & 16.3 (0.4) \\
                   &                       &L& 18.1 (0.4) & 9.3 (0.7) & 14.6 (0.8) & 17.0 (0.5) & 17.2 (0.4) & 15.0 (1.0) & 17.1 (0.5) & 17.2 (0.5) \\\cmidrule{2-11}
                   & \multirow{2}{*}{50\%} &S& 17.1 (0.4) & 13.3 (0.5) & 14.4 (0.8) & 11.5 (1.1) & 16.2 (0.5) & 15.1 (0.7) & 10.5 (1.4) & 16.3 (0.4) \\ 
                   &                       &L& 18.0 (0.4) & 13.8 (0.6) & 15.2 (0.9) & 13.8 (1.1) & 17.2 (0.4) & 15.9 (0.7) & 14.4 (1.2) & 17.2 (0.4) \\ \cmidrule{2-11}
                   & \multirow{2}{*}{70\%} &S& 17.1 (0.4) & 13.2 (0.4) & 13.4 (0.8) & 11.7 (0.9) & 15.5 (0.5) & 14.1 (0.9) & 12.3 (0.8) & 15.5 (0.5) \\ 
                   &                       &L& 18.0 (0.4) & 14.0 (0.4) & 15.2 (0.8) & 12.0 (0.9) & 16.6 (0.5) & 16.0 (0.7) & 12.7 (0.8) & 16.6 (0.5) \\ \hline
\multirow{8}{*}{T} & \multirow{2}{*}{0\%}  &S& 19.4 (0.4) & 11.4 (0.8) & 16.1 (0.9) & 18.6 (0.4) & 18.8 (0.4) & 17.0 (0.9) & 18.7 (0.4) & 18.8 (0.4) \\ 
                   &                       &L& 20.3 (0.3) & 10.1 (0.8) & 15.4 (1.1) & 19.6 (0.4) & 19.7 (0.4) & 16.4 (1.3) & 19.7 (0.4) & 19.8 (0.4) \\ \cmidrule{2-11}
                   & \multirow{2}{*}{20\%} &S& 19.4 (0.3) & 11.9 (0.8) & 16.4 (0.9) & 18.5 (0.4) & 18.7 (0.4) & 17.3 (0.8) & 18.7 (0.4) & 18.8 (0.4) \\ 
                   &                       &L& 20.3 (0.4) & 10.6 (0.8) & 16.1 (1.1) & 19.6 (0.4) & 19.7 (0.4) & 17.0 (1.1) & 19.7 (0.3) & 19.8 (0.4) \\ \cmidrule{2-11}
                   & \multirow{2}{*}{50\%} &S& 19.4 (0.4) & 16.0 (0.5) & 16.7 (0.9) & 13.6 (1.2) & 18.7 (0.4) & 17.7 (0.7) & 14.0 (1.5) & 18.8 (0.4) \\ 
                   &                       &L& 20.3 (0.4) & 16.5 (0.6) & 17.5 (1.0) & 16.5 (1.1) & 19.7 (0.4) & 18.4 (0.8) & 17.8 (0.9) & 19.8 (0.4) \\ \cmidrule{2-11}
                   & \multirow{2}{*}{70\%} &S& 19.4 (0.4) & 15.7 (0.4) & 15.8 (1.0) & 14.2 (0.8) & 18.1 (0.5) & 16.7 (0.9) & 15.1 (0.7) & 18.1 (0.4) \\ 
                   &                       &L& 20.3 (0.4) & 16.5 (0.4) & 17.7 (0.8) & 14.6 (0.9) & 19.2 (0.5) & 18.6 (0.7) & 15.4 (0.8) & 19.2 (0.4) \\ \hline
\end{tabular}
\end{sidewaystable}

\section{Application}\label{App}
High blood pressure increases the risk of cardiovascular disease (CVD) events, which are leading causes of death worldwide\cite{b42}. About half of the U.S. adults have hypertension, which accounts for nearly \$131 billion healthcare expenditure per year\cite{b51}. The Systolic Blood Pressure Intervention Trial (SPRINT) found the intensive treatment that targets a systolic blood pressure (SBP) of $<$120 mm Hg reduced CVD events and all-cause mortality compared to the standard treatment with a SBP goal of $<$140 mm Hg in high CVD risk patients\cite{b46}. However, these findings may be insufficient to inform healthcare policy making due to the lack of economic evidence of the intervention. Bress et al.\cite{b15} projected lifetime costs and effects of the treatment in SPRINT using a microsimulation model and conducted a CE analysis that showed the intensive treatment is cost-effective at a population level. Our analysis extends previous findings by studying the CE of the intensive treatment at a patient level. We adopted the same simulation model in Bress et al.\cite{b15} and applied our proposed method to estimate the most cost-effective treatment rule that is tailored to subjects' characteristics. The model generated 10,000 patients by randomly sampling the SPRINT participants and projected 15-year overall survival and total cost under the assumption that the adherence and treatment effects reduce after 5 years. Our NMB outcome is a composite of restricted survival time and the total cost of hypertension treatment, CVD event hospitalizations, chronic CVD treatment, serious adverse events, and background healthcare. A 3\% discount rate was used for cost. The analysis was repeated for two WTP thresholds recommended by the American Heart Association, \$50K and \$100K per life-year\cite{b61}.

We identified the optimal CE-ITR using the top-performing CRF-AIPW-P method indicated by our simulation study. Since the only source of censoring in our data is administrative censoring, all subjects had complete outcomes data up to 15 years so we used a censoring weight of one in the analysis. We used the same parametric models in the simulation study to estimate the counterfactual restricted survival time and costs. For each outcome model, we specified both the main effect terms and the effect modification terms with the same 17 baseline covariates that pre-determined based on background knowledge. These estimated outcomes were then used in the AIPW-P estimator. We conducted 10-fold cross-validation to obtain the out-of-sample estimates of the optimal rules and the mean NMBs. For each of the 10 iterations, we used 9/10 of the data to estimate the classification weights and train a conditional forest and used the rest 1/10 data as the test data for making predictions. To capture the uncertainty of our estimates, we computed the 95\% confidence intervals with 1000 bootstrap samples\cite{b64}, in which the lower and upper bounds are the 2.5\textsuperscript{th} and the 97.5\textsuperscript{th} percentiles of the bootstrap sampling distribution.

Table \ref{t:DataAnalysis} shows that the average treatment effects are positive across both WTP thresholds, which are the evidence for a conventional CE analysis to recommend the intensive treatment for the entire study cohort\cite{b15}. However, with the consideration of individual heterogeneity, our method reported that the intensive treatment is cost-effective for 69\% of the patients if they are willing to afford up to \$50K for per life-year gain, and the proportion increased to 74\% if the affordability goes up to \$100K per life-year. Besides, our evaluation of the mean NMBs under each treatment rule shows that the personalized regime resulted in larger mean NMBs than the uniform rules for both WTP thresholds. This informs healthcare decision makers that individualized treatment rules should be used to maximize the utility of clinical resources for managing hypertension. Our estimates of the average treatment effects reflect the intervention adherence rate in the data and the assumption that treatment effect fades away after the first 5 years of follow-up.

\begin{sidewaystable}
\small
\centering
\caption{The Proportions of Treated and Estimated Mean Outcomes in A SPRINT-Eligible Cohort. Our effectiveness measure is life-years and the follow-up length is 15 years. The rule of assigning all patients to intensive SBP control is the optimal rule from a conventional cost-effectiveness analysis based on the average treatment effect on NMB across the full study population. We conducted 10-fold cross-validation to avoid overfitting and computed the 95\% CI using bootstrapping.}
\label{t:DataAnalysis}
\begin{tabular}{llccc}
\hline
 &        &{All patients assigned to}&{All patients assigned to} &{AIPW-based CE ITR}\\
 & {WTP}  & {intensive SBP control}  & {standard SBP control} & {(95\% CI)} \\
\hline
\multirow{2}{*} {Proportion of Treated} &50K  &1&0&69\% (49\%, 84\%)\\
                                        &100K &1&0&74\% (54\%, 87\%)\\\hline
\multirow{2}{*} {Mean NMB Outcome} &50K  &437212&429848&438166 (4339245, 442638)\\
                                   &100K &1059798&1037198&1060619 (1053006, 1068140)\\\hline
\end{tabular}
\end{sidewaystable}

To further assess the capability of our method in handling censoring under a real world setting, we followed the suggestions from one of the reviewers and added arbitrary censoring to the SPRINT data. We generated the censoring time using a discrete uniform distribution, i.e., $C \sim \mathrm{DUNIF(7-0.05age, 15)}$, which results in a 22\% event rate. Under this set up, we applied the following four methods: CRF-AIPW-P, CRF-AIPW-NP, DT-AIPW-P, and DT-AIPW-NP. The results in Table \ref{t:DataAnalysisSup} are aligned with the results in Table \ref{t:DataAnalysis} and show that CE-ITRs yielded higher mean NMBs than uniform regimes (Treat all or Treat none). In addition, we see that using partitioned NMB weights helps to achieve a more efficient treatment strategy, that is, treating fewer patients but leading to higher mean NMBs. Furthermore, conditional random forests result in better treatment rules than decision trees, especially when partitioned weights are not available, e.g., there is no cost history data. Among all four estimators, CRF-AIPW-P showed the best performance for both WTP thresholds. Figure \ref{f:varimp} displays the variable importance measure of each predictor. Both CRF-AIPW-P (top right) and DT-AIPW-P (bottom right) indicate that whether or not taking statin at baseline and whether or not being a current smoker at baseline are the top two important variables for estimating CE-ITR for both WTP thresholds.

\section{Discussion}\label{sec:diss}
There is a growing interest in evaluating healthcare resource use in the past decade. Many clinical trials that often designed to study the clinical benefit of interventions now also value the collection of economic data. Some researchers also gather cost data by linking patients' clinical information to claims data such as the all-payer claims database; others may derive costs based on national guidelines such as the National Average Drug Acquisition Cost (NADAC). To understand such data with increasing availability and complexity and gain insights on the CE of certain intervention, we proposed a weighted classification approach to identify the optimal CE-ITR from a data-driven perspective. We adopted an advanced random forest implementation in which the forests are constructed with unbiased conditional inference trees. Furthermore, we proposed two partitioned estimators for the NMB-based classification weights in which the incomplete cost data of censored subjects are utilized to improve the estimation performance. To advocate personalized treatment plan, our methods estimate the optimal regime as a function of individual level characteristics for situations where there may be heterogeneity in treatment effects on both effectiveness and cost. We proposed methods under an observational study setting so that real-world challenges such as right censoring and confounding are considered. The direct application of our proposed methods to randomized controlled trials is also warranted by simply replacing the inverse probability treatment weights with randomization probabilities\cite{b38,b48}.

One limitation of our proposed methods is that they may be sensitive to how a follow-up period is partitioned into subintervals. Using oversized subintervals may lose some cost data from censored subjects due to rounding of time in the last subinterval, while tiny intervals may include too few events and induce extreme censoring weights, especially in studies with low event rates. Also, a large number of subintervals lead to a higher computation burden. When applying our methods, one should be aware of two practical challenges in a CE analysis. First, CE analyses require a set of input parameters such as unit cost or discount rate, whose varied specifications may inform different decision rules; so, sensitivity analysis should be implemented to assess the uncertainty in parameters\cite{b55}. Other input parameters such as event rates or treatment benefits are usually estimated from clinical trials or even observational studies, which may induce selection and confounding bias. So, it is also essential to review and assess the quality of original analysis. Second, due to the limited follow-up in study samples, within-trial parameter estimates are often projected to a lifetime horizon with certain assumptions. For example, one may assume the treatment effect estimated from a 3-year study remains the same for a hundred-year projection, which may be implausible and yield unreasonable results. High-quality CE analysis should be conducted and repeated with multiple justifiable input parameters that reflect the most realistic scenarios. In the meantime, the uncertainty associated with the projection should be evaluated.

We followed the common practice in CE analysis alongside trials and adopted a clinical endpoint as the effectiveness outcome; while other CE analyses that are interested in both the quantity and quality of patients' life may use quality-adjusted-life-year (QALY) as an alternative. QALY is measured as the area under the curve of a quality-of-life function that defined by time and health state utilities. One challenge pointed out by Gelber et al.\cite{b58} in estimating QALYs is the induced informative censoring. Since our methods already account for the same issue for cost, they should be easily extended to situations where NMBs are defined using QALYs. Moreover, in clinical practice, it has become increasingly attractive to extend ITR studies to identify the optimal dynamic treatment regime (DTR) that allows the treatment to change according to a patient's evolving disease status. Several methods have been developed to address various challenges in this extension when only a single health outcome is of interest\cite{b102,b103}. Recently, classification-based direct estimation methods have also demonstrated their contributions to the estimation of optimal DTR\cite{b104}. All these pioneer work provides a solid foundation for our future work on extending the current proposal to estimate the most cost-effective DTR. At each stage, the optimal treatment decision is identified based on patient-specific knowledge of historical treatment benefits and costs; so, the most cost-effective DTR is determined by adjusting the current intervention plan to updated information over time. The development of DTR is extraordinarily beneficial to patients with chronic diseases\cite{b59} as long-term medication intake may increase the risk of drug toxicity and resistance. Also, the extension of DTR with a CE perception helps to control the potentially high medical expenses induced by long-term care and optimize resource allocation in chronic diseases.

\begin{acks}
This work was directly supported by R01 HL139837 from the National Heart, Lung, and Blood Institute (NHLBI), Bethesda, MD., and the Utah Center for Clinical and Translational Science. Dr. Bellows is supported by K01 HL140170 from the National Heart, Lung, and Blood Institute.
\end{acks}

\begin{dci}
The author(s) declared no potential conflicts of interest with respect to the research, authorship, and/or publication of this article.
\end{dci}



\begin{sm}

\subsection*{Part I. Additional Simulation Results}
Additional simulation results for willingness-to-pay = \$100K that referenced in the Simulation Results Section.

\begin{sidewaystable}
\small
\centering
\caption{Comparison of Classification Accuracies (\%) of Estimated Optimal ITRs. ``EM=TM" indicates the presence of effect modification on both survival time and cost; ``EM=T" indicates the presence of effect modification on survival time only. ``CR" is short for ``censoring rate". ``HTE" indicates the amount of effect modification on the hazard ratio scale for survival time: ``HTE=S" for small effect modification and ``HTE=L" for large effect modification. Willingness-to-pay = \$100K.}
\label{t:CCR100K}
\begin{tabular}{lllccccccc}
\hline
&&&&&&&&&\\
EM& CR & HTE & \multicolumn{1}{c}{Reg-naive} & \multicolumn{1}{c}{DT-AIPW-NP} & \multicolumn{1}{c}{DT-IPW-P}& \multicolumn{1}{c}{DT-AIPW-P} & \multicolumn{1}{c}{CRF-AIPW-NP} & \multicolumn{1}{c}{CRF-IPW-P}& \multicolumn{1}{c}{CRF-AIPW-P}\\
\hline
\multirow{8}{*}{TM}& \multirow{2}{*}{0\%}  &S& 73.4 (2.3) & 91.7 (1.8) & 91.3 (1.4) & 93.0 (1.2) & 93.6 (1.0) & 92.6 (1.0) & 93.7 (1.0) \\ 
                   &                       &L& 67.5 (3.2) & 88.9 (2.8) & 92.2 (1.3) & 93.8 (1.0) & 92.3 (2.0) & 93.3 (0.9) & 94.3 (0.9) \\\cmidrule{2-10}
                   & \multirow{2}{*}{20\%} &S& 75.0 (2.0) & 92.1 (1.7) & 91.3 (1.4) & 92.9 (1.2) & 93.7 (1.0) & 92.5 (1.1) & 93.7 (1.0) \\
                   &                       &L& 71.1 (2.8) & 90.4 (2.3) & 92.3 (1.3) & 93.7 (1.1) & 93.4 (1.5) & 93.3 (1.0) & 94.2 (0.9) \\ \cmidrule{2-10}
                   & \multirow{2}{*}{50\%} &S& 78.7 (1.4) & 92.4 (1.4) & 90.1 (1.7) & 92.8 (1.2) & 93.2 (1.0) & 91.8 (1.2) & 93.3 (1.0) \\ 
                   &                       &L& 79.2 (1.4) & 93.4 (1.4) & 91.8 (1.4) & 93.6 (1.1) & 94.2 (0.9) & 92.9 (1.1) & 94.1 (0.9) \\\cmidrule{2-10}
                   & \multirow{2}{*}{70\%} &S& 78.4 (1.4) & 77.1 (4.8) & 80.6 (2.7) & 89.3 (1.8) & 81.4 (4.7) & 83.5 (2.3) & 89.7 (1.5) \\
                   &                       &L& 79.0 (1.3) & 85.7 (3.4) & 80.5 (2.9) & 91.5 (1.5) & 89.5 (2.3) & 84.1 (2.6) & 91.4 (1.5) \\\hline
\multirow{8}{*}{T} & \multirow{2}{*}{0\%}  &S& 73.7 (2.2) & 92.3 (1.6) & 91.4 (1.4) & 92.9 (1.2) & 93.7 (1.0) & 92.5 (1.2) & 93.7 (1.0) \\ 
                   &                       &L& 68.6 (3.1) & 89.7 (2.8) & 92.2 (1.3) & 93.7 (1.0) & 93.1 (1.6) & 93.3 (1.0) & 94.3 (0.8) \\ \cmidrule{2-10}
                   & \multirow{2}{*}{20\%} &S& 75.1 (1.9) & 92.6 (1.4) & 91.4 (1.4) & 92.9 (1.1) & 93.7 (1.0) & 92.6 (1.1) & 93.6 (1.0) \\ 
                   &                       &L& 71.9 (2.6) & 91.2 (2.1) & 92.3 (1.3) & 93.6 (1.0) & 93.8 (1.3) & 93.3 (1.0) & 94.2 (0.9) \\ \cmidrule{2-10}
                   & \multirow{2}{*}{50\%} &S& 78.9 (1.4) & 92.4 (1.4) & 90.5 (1.6) & 92.7 (1.2) & 93.3 (1.0) & 91.9 (1.3) & 93.3 (1.0) \\
                   &                       &L& 79.3 (1.4) & 93.4 (1.3) & 92.0 (1.4) & 93.5 (1.1) & 94.1 (0.9) & 93.0 (1.1) & 93.9 (1.0) \\ \cmidrule{2-10}
                   & \multirow{2}{*}{70\%} &S& 78.3 (1.4) & 77.5 (4.9) & 81.0 (2.8) & 89.4 (1.7) & 81.6 (4.5) & 83.9 (2.3) & 89.5 (1.7) \\
                   &                       &L& 78.8 (1.3) & 85.8 (3.4) & 80.7 (2.9) & 91.5 (1.5) & 89.5 (2.3) & 84.4 (2.4) & 91.4 (1.6) \\ \hline
\end{tabular}
\end{sidewaystable}

\begin{sidewaystable}
\fontsize{8}{12}\selectfont
\centering
\caption{Comparison of Mean Outcomes under Different Estimated Optimal ITRs. The indications of ``EM", ``CR", ``WTP", and ``HTE" are the same as those in Table 1. The $g^{\mathrm{opt}}$ column represents the mean NMB (in the unit of 10,000) under the true optimal regime, and the other seven columns on the right display the mean NMB under the corresponding estimated regime. Willingness-to-pay = \$100K.}
\label{t:EMO100K}
\begin{tabular}{lllcccccccc}
\hline
&&&&&&&&&&\\
EM& CR & HTE &\multicolumn{1}{c}{$\mathbb{E}[Y^{g^{\mathrm{opt}}}]$}& \multicolumn{1}{c}{Reg-naive} & \multicolumn{1}{c}{DT-AIPW-NP} & \multicolumn{1}{c}{DT-IPW-P}& \multicolumn{1}{c}{DT-AIPW-P} & \multicolumn{1}{c}{CRF-AIPW-NP} & \multicolumn{1}{c}{CRF-IPW-P}& \multicolumn{1}{c}{CRF-AIPW-P}\\
\hline
\multirow{8}{*}{TM}& \multirow{2}{*}{0\%}  &S&  98.8 (1.2) & 79.9 (2.5) & 95.9 (2.0)   & 96.2 (1.4) & 97.3 (1.3) & 97.6 (1.4) & 96.7 (1.4) & 97.7 (1.4) \\  
                   &                       &L& 103.6 (1.1) & 76.8 (3.6) & 96.9 (3.2)  & 101.0 (1.4) & 102.1 (1.2) & 100.4 (2.4) & 101.5 (1.3) & 102.5 (1.2) \\ \cmidrule{2-11}
                   & \multirow{2}{*}{20\%} &S&  98.7 (1.2) & 81.5 (2.2) & 96.3 (1.9)   & 96.1 (1.4) & 97.2 (1.3) & 97.6 (1.3) & 96.7 (1.3) & 97.7 (1.2) \\  
                   &                       &L& 103.6 (1.2) & 80.7 (3.1) & 98.6 (2.6)  & 101.0 (1.3) & 102.1 (1.3) & 101.8 (1.8) & 101.7 (1.2) & 102.7 (1.2) \\  \cmidrule{2-11}
                   & \multirow{2}{*}{50\%} &S&  98.7 (1.3) & 85.4 (1.6) & 96.8 (1.6)  & 94.9 (1.7) & 97.1 (1.4) & 97.5 (1.2) & 96.1 (1.3) & 97.5 (1.2) \\ 
                   &                       &L& 103.6 (1.2) & 89.2 (1.6) & 101.8 (1.5) & 100.5 (1.6) & 102.0 (1.3) & 102.4 (1.3) & 101.2 (1.3) & 102.4 (1.2) \\  \cmidrule{2-11}
                   & \multirow{2}{*}{70\%} &S&  98.7 (1.3) & 85.1 (1.6) & 84.4 (4.9)  & 87.7 (2.6) & 94.5 (1.8) & 89.0 (4.4) & 90.3 (2.1) & 94.8 (1.5) \\ 
                   &                       &L& 103.5 (1.2) & 88.9 (1.5) & 95.3 (3.6)  & 90.6 (2.8) & 100.1 (1.6) & 99.2 (2.3) & 94.2 (2.6) & 100.2 (1.6) \\  \hline
\multirow{8}{*}{T} & \multirow{2}{*}{0\%}  &S& 101.3 (1.3) & 82.5 (2.5) & 99.1 (1.9)  & 98.7 (1.4) & 99.8 (1.3) & 100.1 (1.3) & 99.2 (1.4) & 100.2 (1.3) \\
                   &                       &L& 106.1 (1.2) & 79.6 (3.7) & 100.0 (3.3) & 103.6 (1.4) & 104.7 (1.2) & 103.7 (1.9) & 104.1 (1.3) & 105.1 (1.2) \\ \cmidrule{2-11}
                   & \multirow{2}{*}{20\%} &S& 101.2 (1.2) & 84.0 (2.3) & 99.4 (1.7)  & 98.7 (1.4) & 99.7 (1.3) & 100.4 (1.4) & 99.4 (1.4) & 100.4 (1.4) \\ 
                   &                       &L& 106.2 (1.2) & 83.6 (3.2) & 101.7 (2.5) & 103.6 (1.3) & 104.6 (1.3) & 104.5 (1.6) & 104.1 (1.3) & 105.0 (1.3) \\\cmidrule{2-11}
                   & \multirow{2}{*}{50\%} &S& 101.2 (1.3) & 88.1 (1.7) & 99.4 (1.5)  & 97.8 (1.6) & 99.6 (1.4) & 100.2 (1.3) & 98.8 (1.5) & 100.1 (1.3) \\ 
                   &                       &L& 106.1 (1.2) & 92.0 (1.7) & 104.4 (1.5) & 103.2 (1.5) & 104.5 (1.3) & 105.0 (1.2) & 103.9 (1.3) & 105.0 (1.2) \\ \cmidrule{2-11}
                   & \multirow{2}{*}{70\%} &S& 101.2 (1.3) & 87.5 (1.6) & 86.9 (5.3) & 90.4 (2.6) & 97.1 (1.8) & 91.4 (4.4) & 92.9 (2.2) & 97.1 (1.6)\\
                   &                       &L& 106.1 (1.2) & 91.4 (1.6) & 97.9 (3.8) & 93.2 (2.9) & 102.6 (1.6) & 101.9 (2.4) & 97.0 (2.3) & 102.8 (1.6) \\ \hline
\end{tabular}
\end{sidewaystable}

\subsection*{Part II. Additional Figures}
Additional decision boundary figures (Figures 3-8) that referenced in the Simulation Results Section.

\begin{sidewaysfigure}
\centering
\centerline{\includegraphics[scale=0.43]{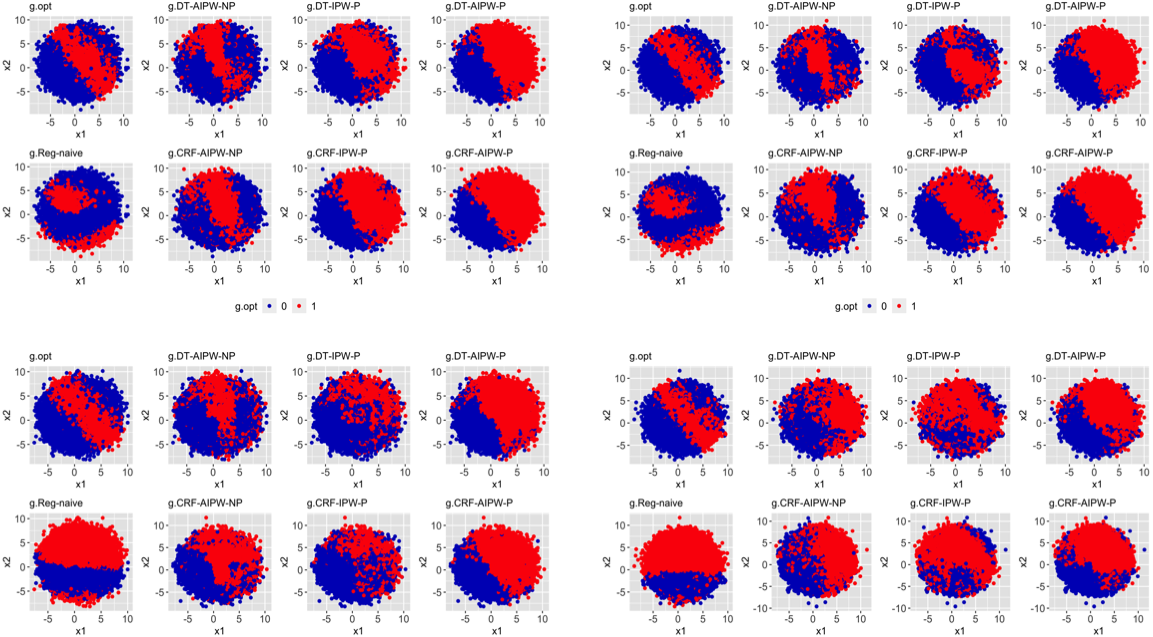}}
\caption{Comparison of Decision Boundaries of Estimated Optimal ITRs. Presence of effect modification on both survival time and cost, and only the outcome models are misspecified. WTP = \$50K and small HTE. The four panels are labeled as follows. Top left:, CR=0\%; Top right:, CR=20\%; Bottom left: CR=50\%; Bottom right: CR=70\%. Each panel contains eight figures: the very first figure shows the true decision boundary (gold standard), and the rest seven display the estimated decision boundaries that are given by different methods.}
\label{f:SimMap}
\end{sidewaysfigure}

\begin{sidewaysfigure}
\centering
\centerline{\includegraphics[scale=0.43]{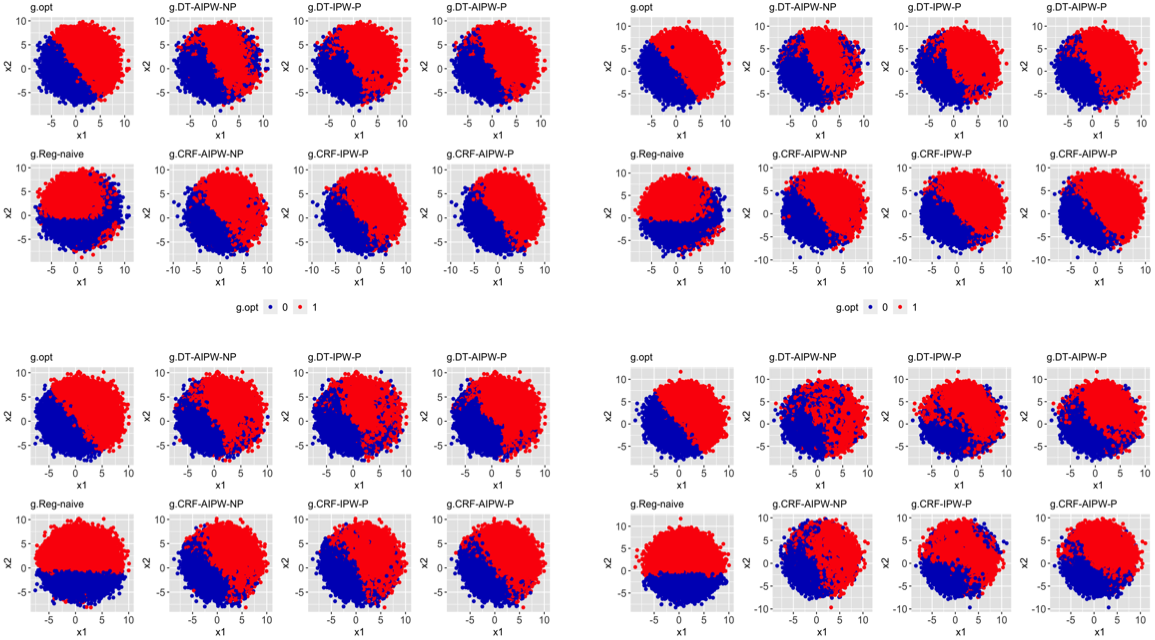}}
\caption{Comparison of Decision Boundaries of Estimated Optimal ITRs. Presence of effect modification on both survival time and cost, and only the outcome models are misspecified. WTP = \$100K and small HTE. The four panels are labeled as follows. Top left:, CR=0\%; Top right:, CR=20\%; Bottom left: CR=50\%; Bottom right: CR=70\%. Each panel contains eight figures: the very first figure shows the true decision boundary (gold standard), and the rest seven display the estimated decision boundaries that are given by different methods.}
\end{sidewaysfigure}

\begin{sidewaysfigure}
\centering
\centerline{\includegraphics[scale=0.43]{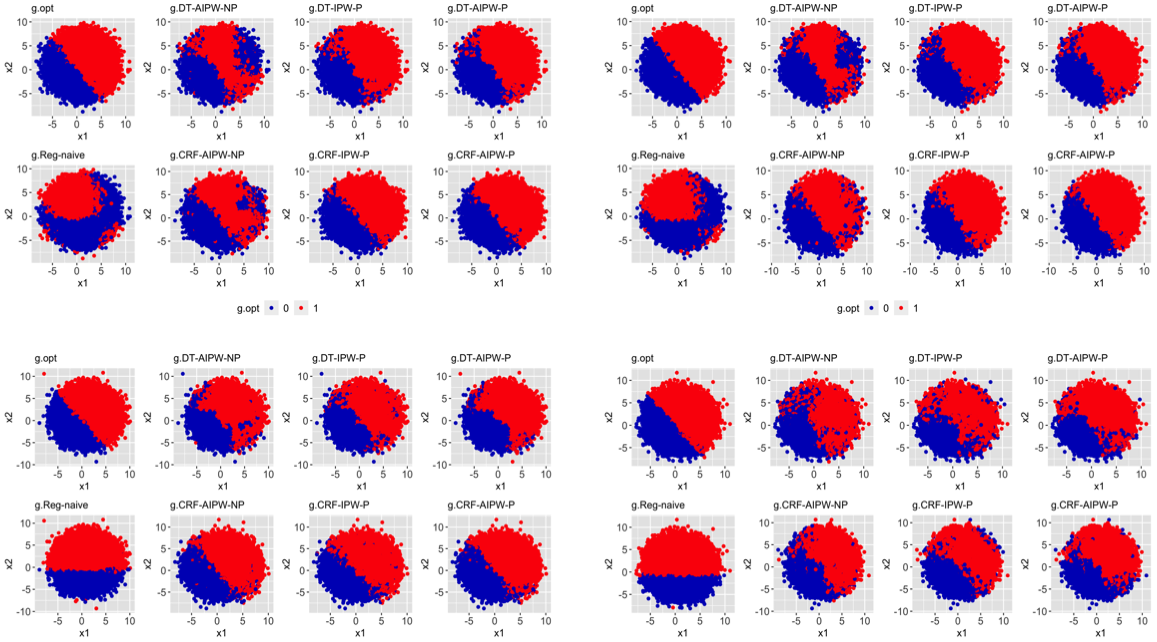}}
\caption{Comparison of Decision Boundaries of Estimated Optimal ITRs. Presence of effect modification on both survival time and cost, and only the outcome models are misspecified. WTP = \$100K and large HTE. The four panels are labeled as follows. Top left:, CR=0\%; Top right:, CR=20\%; Bottom left: CR=50\%; Bottom right: CR=70\%. Each panel contains eight figures: the very first figure shows the true decision boundary (gold standard), and the rest seven display the estimated decision boundaries that are given by different methods.}
\end{sidewaysfigure}

\begin{sidewaysfigure}
\centering
\centerline{\includegraphics[scale=0.43]{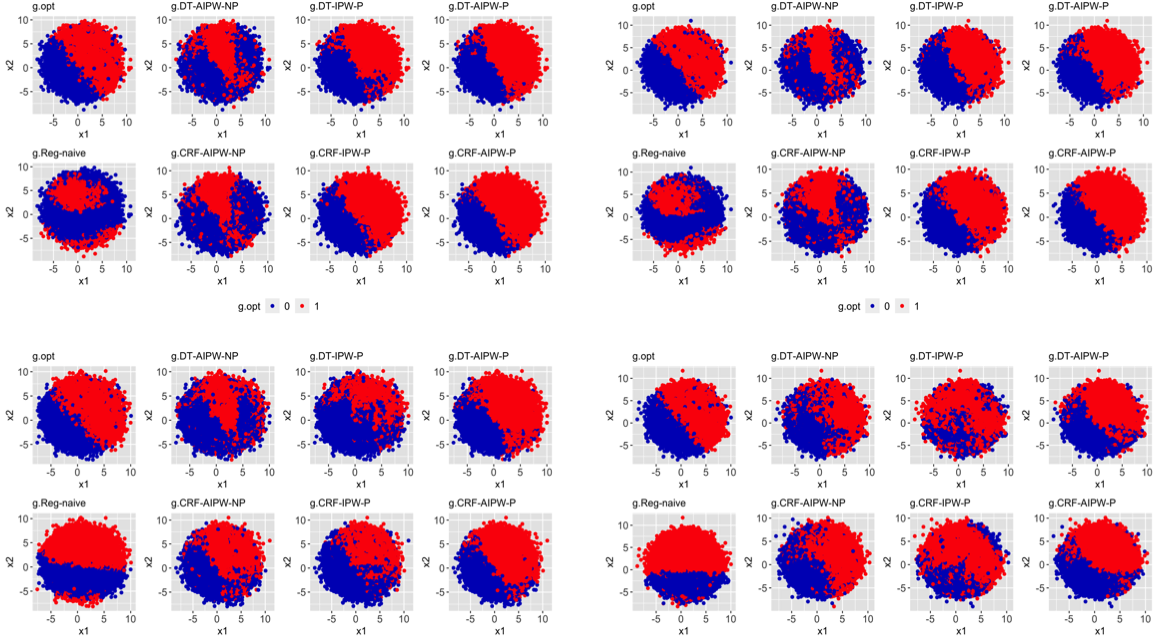}}
\caption{Comparison of the Decision Boundaries of the Estimated Optimal ITRs. Presence of effect modification on survival time but not on cost, and only the outcome models are misspecified. WTP = \$50K and small HTE. The four panels are labeled as follows. Top left:, CR=0\%; Top right:, CR=20\%; Bottom left: CR=50\%; Bottom right: CR=70\%. Each panel contains eight figures: the very first figure shows the true decision boundary (gold standard), and the rest seven display the estimated decision boundaries that are given by different methods.}
\end{sidewaysfigure}

\begin{sidewaysfigure}
\centering
\centerline{\includegraphics[scale=0.43]{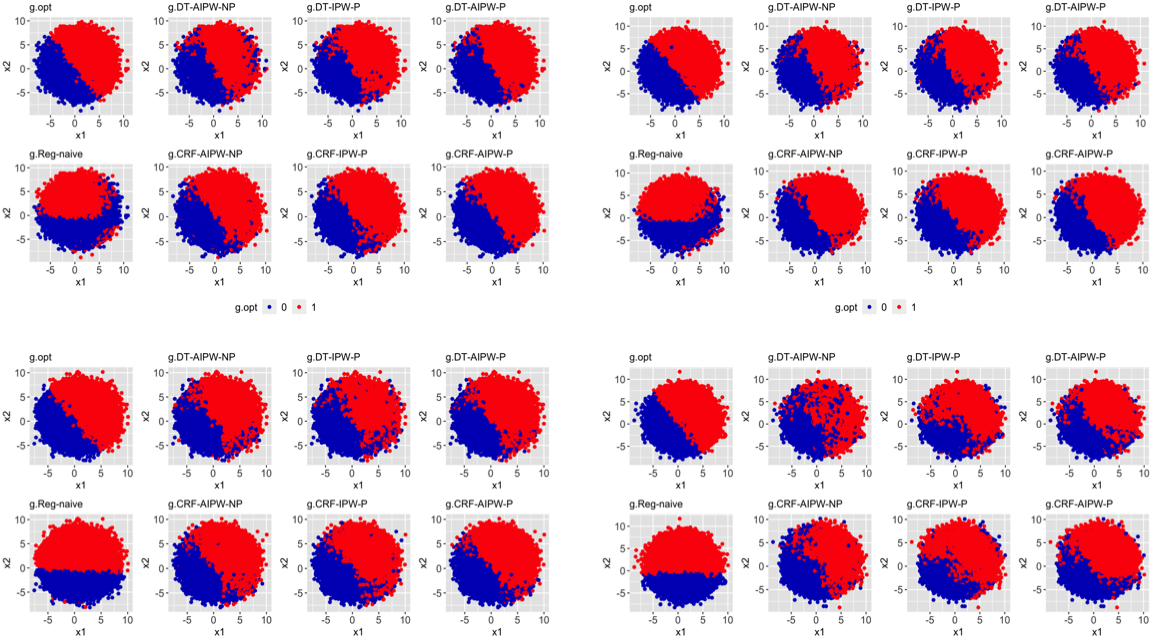}}
\caption{Comparison of the Decision Boundaries of the Estimated Optimal ITRs. Presence of effect modification on survival time but not on cost, and only the outcome models are misspecified. WTP = \$100K and small HTE. The four panels are labeled as follows. Top left:, CR=0\%; Top right:, CR=20\%; Bottom left: CR=50\%; Bottom right: CR=70\%. Each panel contains eight figures: the very first figure shows the true decision boundary (gold standard), and the rest seven display the estimated decision boundaries that are given by different methods.}
\end{sidewaysfigure}

\begin{sidewaysfigure}
\centering
\centerline{\includegraphics[scale=0.43]{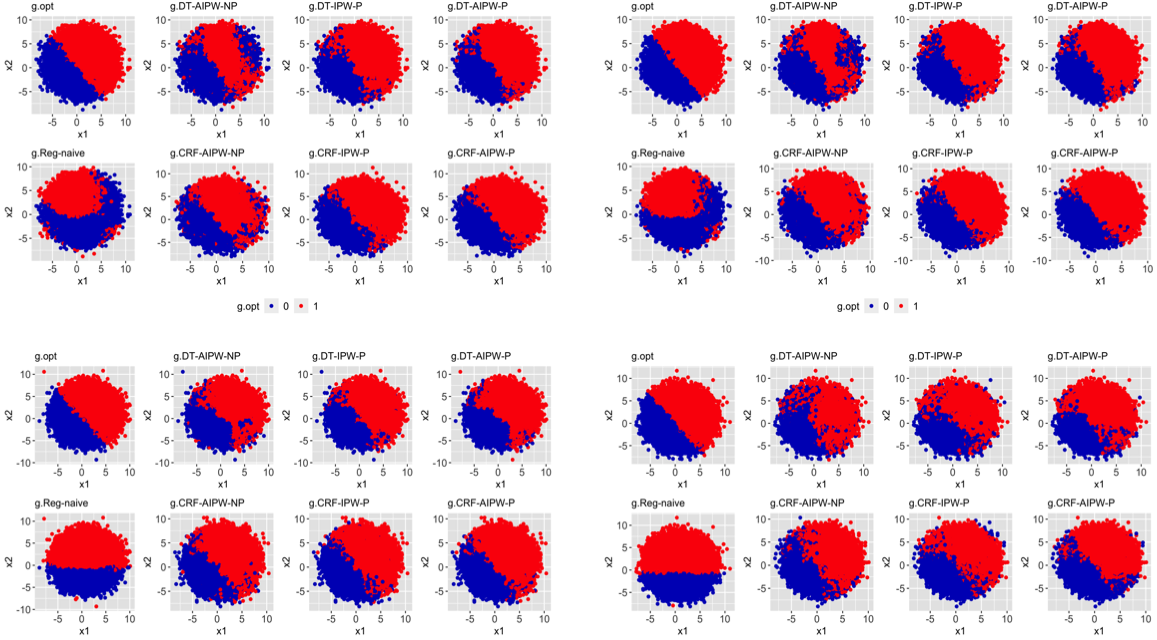}}
\caption{Comparison of the Decision Boundaries of the Estimated Optimal ITRs. Presence of effect modification on survival time but not on cost, and only the outcome models are misspecified. WTP = \$100K and large HTE. The four panels are labeled as follows. Top left:, CR=0\%; Top right:, CR=20\%; Bottom left: CR=50\%; Bottom right: CR=70\%. Each panel contains eight figures: the very first figure shows the true decision boundary (gold standard), and the rest seven display the estimated decision boundaries that are given by different methods.}
\end{sidewaysfigure}

\subsection*{Part III. Additional Real Data Analyses}
\label{sec:sm}
Additional data analysis results on the SPRINT data that requested by reviewers.

\begin{table}
\small
\centering
\caption{Various Blood Pressure Control Strategies for A SPRINT-Eligible Cohort. The event rate is 22\%. Treat all: assign the intensive therapy to everyone, Treat none: assign the standard therapy to everyone, CRF-AIPW-P: CE-ITR is estimated using a conditional forest with AIPW partitioned weights, CRF-AIPW-NP: a conditional forest with AIPW non-partitioned weights, DT-AIPW-P: a decision tree with AIPW partitioned weights, DT-AIPW-NP: a decision tree with AIPW non-partitioned weights.}
\label{t:DataAnalysisSup}
\begin{tabular}{lcccc}
\hline
 &\multicolumn{2}{c}{WTP=50K} &\multicolumn{2}{c}{WTP=100K}\\\hline
 & \% Treated & Mean NMB  & \% Treated & Mean NMB \\\hline
Treat all &100\% &379397 &100\% &943004\\
Treat none &0\% &377622 &0\% &931400\\
CRF-AIPW-P&49.7\% &387246 &50.1\% &956029\\
CRF-AIPW-NP&31.1\% &382403 &39.0\% &943435\\
DT-AIPW-P&53.7\% &387146 &53.7\% &956001\\
DT-AIPW-NP&62.5\% &378682 &62.5\% &938548\\\hline
\end{tabular}
\end{table}

\begin{figure}
\centering
\centerline{\includegraphics[scale=0.13]{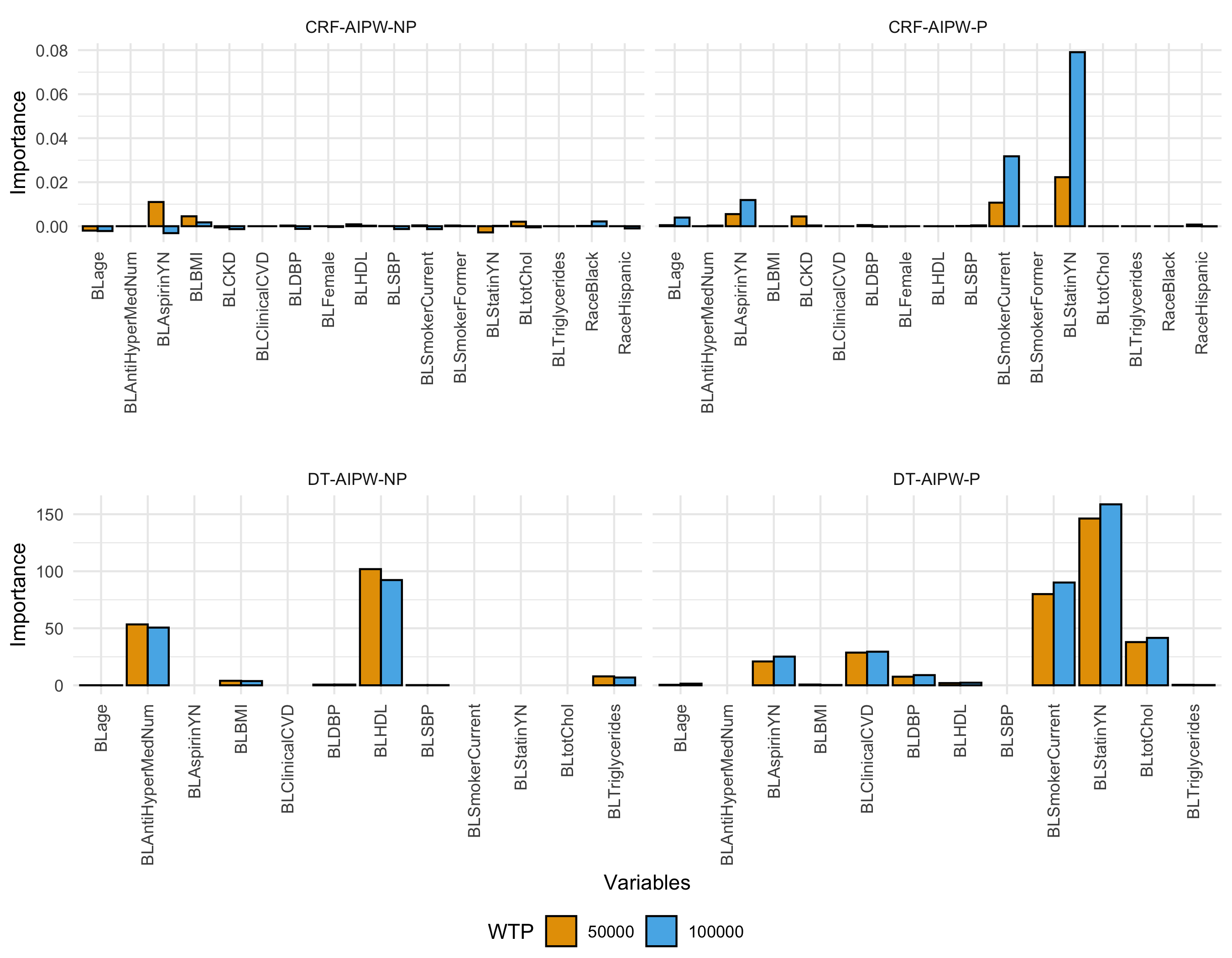}}
\caption{Variable Importance of Baseline Covariates for Estimating Cost-effective Individualized Treatment Rules in SPRINT.}
\label{f:varimp}
\end{figure}

\subsection*{Part IV. Analysis Code}
Data and R code are available at \url{https://github.com/CrystalXuR/EfficientCEITR}.
\end{sm}
\end{document}